\documentclass[11pt,a4paper,final]{article}
\usepackage{ifpdf}               
\usepackage{a4wide}             
\usepackage{amssymb}            
\usepackage{amsmath}          
\usepackage{slashed}            
\usepackage[usenames,dvipsnames]{color}
\usepackage[latin1]{inputenc}    
\usepackage[notref,notcite]{showkeys} 
\ifpdf
\usepackage[pdftex]{graphicx}
\usepackage[pdftex,unicode,implicit]{hyperref}
\hypersetup{%
  pdftitle    = {},
  pdfkeywords = {},
  pdfauthor   = {},
  pdfcreator  = {pdf\LaTeXe\ with package \flqq hyperref\frqq},
  pdfproducer = {pdf\LaTeXe\ with package \flqq hyperref\frqq},
  pdfpagemode = None,  
  pdffitwindow= true,  
  unicode     = true,
  plainpages  = false,
  colorlinks  = true,  
  citecolor   = black,
  urlcolor    = black,
  linkcolor   = black
}

\else
  \usepackage[dvips]{graphicx}
  \usepackage[unicode,implicit]{hyperref}

\fi
\usepackage{tikz}
\newcommand{\FPAUO}[2]{
\tikz[scale=.13,
         Uniovi/.style={color=gray, fill=gray}
 ] {
 \fill[Uniovi] (0,0) circle (10);
 \fill[white] (0,7) circle (1.5);
 \draw[Uniovi] (-2,7.5) rectangle (2,5.5);
 \fill[white] (-0.3,6.6) rectangle (0.3,0);   
 \fill[white] ( -0.9,6.2) rectangle (.9 ,5.6);
 \fill[white] (-1.4, 5.2) rectangle (1.4, 4.6);
 \fill[white] (0,0) ellipse (3.5 and 4);
 \fill[Uniovi] (-2.5,0.3) rectangle (2.5,-0.3);
 \fill[Uniovi] (-2,2.3) rectangle (2,1.7);
 \fill[Uniovi] (-2,-2.3) rectangle (2,-1.7);
 \fill[white] (-4.5,5.5) rectangle (-2.7,4.9);
 \fill[white] (-3.9,6.1) rectangle (-3.3,4.3);
 \fill[white] (4.5,5.5) rectangle (2.7,4.9);
 \fill[white] (3.9,6.1) rectangle (3.3,4.3);
 \foreach \x in { 0,..., 3 }
   \foreach \y in { 0,...,\x}
    {
     \fill[white] (-6-\x*0.7+\y*1.4,3.5-\x *1.97) -- (-5.6-\x*0.7+\y*1.4,2.4-\x *1.97) -- (-6.4-\x*0.7+\y*1.4,2.4-\x *1.97) -- cycle;
     \fill[white] (6-\x*0.7+\y*1.4,3.5-\x *1.97) -- (5.6-\x*0.7+\y*1.4,2.4-\x *1.97) -- (6.4-\x*0.7+\y*1.4,2.4-\x *1.97) -- cycle;
   };
 \draw (0,-6) node[
                               text centered, 
                               color=white, 
                               font={\fontsize{8}{4}\sffamily\selectfont}
                             ] {FPAUO-#1/#2};
}} 
\newcommand{\change}[1]{#1}
\begin{document}
~\vspace{-4cm}\begin{flushright}
\small
\FPAUO{11}{01}\\
IFT-UAM/CSIC-11-37\\
September 20, 2011
\normalsize
\end{flushright}
\begin{center}
{\Large\bf On supersymmetric Lorentzian Einstein-Weyl spaces}\\[.5cm]
{\bf Patrick~Meessen}$^{\aleph}$,
{\bf Tom\'as Ort\'{\i}n}$^{\sharp}$ and 
{\bf Alberto Palomo-Lozano}$^{\sharp}$\\[.2cm]
$^{\aleph}$ {\em HEP Theory Group, Departamento de F\'{\i}sica, Universidad de Oviedo\\ 
        Avda.~Calvo Sotelo s/n, E-33007 Oviedo, Spain.}\\[.2cm]
$^{\sharp}$ {\em Instituto de F\'{\i}sica Te\'orica UAM/CSIC, C/ Nicol\'as Cabrera 13-15\\
        Ciudad Universitaria de Cantoblanco, E-28049 Madrid, Spain.} 
\\[4ex]

{\small In loving memory of Cte. Lozano Cid}\\
\begin{quote} 
{\small \textsc{Abstract.}  
\change{
   We consider weighted parallel spinors in Lorentzian Weyl geometry in arbitrary dimensions,
   choosing the weight such that the integrability condition for the existence of
   such a spinor, implies the geometry to be Einstein-Weyl. We then use techniques
   developed for the classification of supersymmetric solutions to supergravity theories
   to characterise  those Lorentzian EW geometries that allow for a weighted parallel spinor,
   calling the resulting geometries supersymmetric. The overall result is that they are either
   conformally related to ordinary geometries admitting parallel spinors (w.r.t.~the Levi-Civit\`a connection) 
   or are conformally related to certain Kundt spacetime. 
   A full characterisation is obtained for the 4 and 6 dimensional cases.
}
%
}
\end{quote}
\end{center}

Over the last decades, spinorial fields parallelised by some \change{(generalised)} covariant
derivative\change{ (we shall call such spinorial fields Killing spinors),}\footnote{
  Observe that the concept of Killing spinor \change{(}equation\change{)} in the physics
  literature has a far broader meaning than in the mathematical literature. }
have become a prominent tool in physics as well as mathematics. In
physics, such spinorial fields are usually related to supersymmetry and can be
used to prove the positivity of the energy in physical systems, the stability
of objects that preserve some residual supersymmetry or the
non-renormalisability of the mass-charge relation for the so-called BPS
objects which is of the utmost importance in, for example, String Theory's
microscopic explanation of the entropy of supersymmetric black holes. In
mathematics, one application which also appears frequently in the physics
literature, is the link established by Hitchin between manifolds admitting
parallel spinors and them having a special holonomy group \cite{Hitchin:1974},
but can also be applied to more general settings, such as 
Weyl geometry \cite{art:moroianu1996a,Buchholz:1999a}.\footnote{
 (Global) Spinors also have their applications and can {\em e.g.\/} be used to find
 generalised instanton equations \cite{Detournay:2009mp}.
}
\par
Seeing the importance of such spinors it should not be surprising that in the
last decade techniques were developed to extract the geometric information
contained in the so-called {\em Killing spinor equations} (KSEs)\change{ 
{\em i.e.\/} the equations imposing the parallelity of the spinorial field
under the generalised connection
}.
The first systematic approach was made by Tod in ref.~\cite{Tod:1983pm},
taking leads from earlier work by Gibbons and Hull \cite{Gibbons:1982fy}, who
used the Newman-Penrose spinorial techniques \cite{Penrose:1985jw} to obtain
the supersymmetric solutions to a 4-dimensional supergravity theory usually
referred to as minimal (or \textit{pure}) $N=2$ supergravity.\footnote{
   \change{
   In the supergravity literature it is customary to refer to specific theories
   by indicating the dimension of spacetime, $d$, and the number of minimal 
   spinors, $N$, used to generate the supersymmetry transformations; the
   theory we just mentioned is therefore known as $d=5$ $N=2$ supergravity.
   However, in order not to give too many different meanings to $d$ we will use
   the number $n$ to mean dimensionality of spacetime, and will use the
   non-standard nomenclature ``$N=\sharp$ $n=\sharp$ supergravity''.
   }
}  
In ref.~\cite{Gauntlett:2002nw}, Gauntlett {\em et al.\/} overcame the inherent
4-dimensional restriction of the Newman-Penrose formalism by introducing the
{\em spinor bilinear method} and classifying the supersymmetric solutions of
5-dimensional minimal $N=1$ supergravity. This seminal article was the
starting shot for a period of feverish activity in the supergravity
literature, during which the supersymmetric solutions of the majority of
supergravity theories were characterised, and even more powerful techniques,
such as Gillard {\em et al.\/}'s spinorial geometry method
\cite{Gillard:2004xq}, were developed.
\par
The process of the spinor bilinear characterisation is basically split into
two parts: first, given a rule for the parallel propagation of the spinor in
terms of the relevant supergravity fields, one deduces the most general form
of those fields compatible with the existence of a non-vanishing Killing
spinor; the form of the fields thus obtained is called a supersymmetric field
configuration. Seeing that the KSEs are linear in derivatives and the
equations of motion (EOMs) are of second order, one cannot hope to obtain a
recipe for solutions of the EOMs straight-away, and instead one uses the
supersymmetric configurations as Ans\"atze to find (supersymmetric)
solutions. In this sense, an observation made by Gauntlett {\em et al.\/} in
ref.~\cite{Gauntlett:2002nw} (which was formalised in
ref.~\cite{Bellorin:2005hy}) reduces the amount of work necessary to find the
conditions that a supersymmetric field configuration needs to fulfill, in
order to give rise to a supersymmetric solution.  The basic observation is
that the fact that a solution preserves some supersymmetry means that there
are relations between components of the equations of motion, meaning that
there is a minimal set of independent components of the EOMs that, once
satisfied, implies that all EOMs are satisfied. This observation is in fact
completely general and depends only on a subset of the integrability
conditions for the KSEs under consideration and on the spinorial structure used
\cite{Bellorin:2005hy}.
\par
The first ones to realise that these techniques could be applied outside the
realm of supersymmetry were the authors of ref.~\cite{Grover:2008jr}. They
considered a KSE similar to the one used in 5-dimensional minimal gauged
supergravity, but with a De Sitter-like cosmological constant.\footnote{This
  is in general incompatible with supersymmetry. Sometimes these theories are
  referred to as \textit{fake} supergravities or (f)SUGRA.} As explained
above, the integrability condition of their KSE places a contraint on the
Ricci tensor corresponding to the Einstein's equations of motion, which then
follow automatically from a solution to the KSE. The article goes on
to classify the \emph{timelike} solutions of the constructed theory, which
turn out to show a four-dimensional hyper-K\"ahler torsion (HKT) base space
dependence.
\par
The work we present here follows similar lines, since we also consider a
`novel' KSE (in the sense that such KSE is not related \textit{a priori} to
any supersymmetric setting previously treated), and whose relevance becomes
apparent once one analyses its integrability condition. Our motivation,
however, is different from that of characterisations of solutions to (f)SUGRA
theories. We are interested in classifying Lorentzian Einstein-Weyl spaces of
arbitrary dimension, and the KSE is chosen in such a way that the
integrability condition resembles the geometric constraint for a manifold to
be Einstein-Weyl. The tools we will use for this work are the same ones as used
in the programme of classification of solutions to supergravity theories, and
we will split the problem at hand according to whether they employ a timelike
or null vector field. The characterisation we give is of those EW spaces that
arise from the existence of a \textit{Killing spinor} {\em i.e.\/} a spinor
that fulfills the KSE we propose, and it is in that sense that we refer to
them as supersymmetric.
\par
Section~(\ref{sec:maths}) introduces the spinorial rule, its integrability
condition (which resembles the geometric constraint for Einstein-Weyl spaces)
and a short manipulation on a vector bilinear valid for all dimensions and
cases. Section~(\ref{sec:timelike}) analyses all possible timelike cases,
showing their triviality. Section~(\ref{sec:N1D4}) describes the {\em null}
solutions for the $N=1$, $n=4$ case, while section~(\ref{sec:D6chiral}) treats
the $n=6$ null case and section~(\ref{sec:general}) the remaining
cases. Section~(\ref{sec:conclusions}) recapitulates the work done. Three
appendices are presented at the end for reference and
completeness. Appendix~(\ref{appsec:Weyl}) gives some basic knowledge (by no
means exhaustive) of Weyl geometry and Einstein-Weyl
spaces. Appendix~(\ref{appsec:spinors}) presents the spinorial notation we use
in the article. Appendix~(\ref{appsec:Kundt}) gives the geometrical description
for Kundt waves, as they turn out to be relevant.

\section{Covariant rule and the Einstein-Weyl condition}
\label{sec:maths} 
Consider the following rule for the covariant derivative of some spinor, which
we shall take to be Dirac,
\begin{equation}
  \label{eq:1}
  \nabla_{a}\epsilon \; =\; \textstyle{\frac{4-n}{4}}\ A_{a}\epsilon \; 
+\; \textstyle{1\over 2} \gamma_{ab}A^{b}\epsilon \; ,
\end{equation}
where $n$ is the number of spacetime dimensions and $A$ is just some real
1-form, which at this point is completely unconstrained.  We will call the
solutions $\epsilon$ of this equation Killing spinors and the corresponding
metric and 1-form, a supersymmetric field configuration. Observe that with our
choice of Dirac conjugate, the above rule implies
\begin{equation}
  \label{eq:38}
  \nabla_{a}\overline{\epsilon} \; =\; \textstyle{\frac{4-n}{4}}\
  A_{a}\overline{\epsilon} \; 
-\; \textstyle{1\over 2}A^{b}\ \overline{\epsilon}\gamma_{ab} \; .
\end{equation}
\par
A straightforward calculation of the integrability condition leads to
\begin{equation}
  \label{eq:2}
  \textstyle{1\over 2} \gamma_{a}\slashed{F}\ \epsilon \; =\; \textstyle{1\over 2}\ \mathrm{W}_{(ab)}\gamma^{b}\epsilon \; ,
\end{equation}
where $F\equiv dA$ is called the Faraday tensor and 
\begin{equation}
  \label{eq:14}
  \mathrm{W}_{(ab)} \ =\ \mathtt{R}(g)_{ab}
           \ -\ (n-2) \nabla_{(a}A_{b)}
           \ -\ (n-2)\ A_{a}A_{b}
           \ -\ g_{ab}\ \left[
                        \nabla_{c}A^{c} 
                    \ -\ (n-2)\ A_{c}A^{c}
                   \right] \; ,
\end{equation}
which is readily identified with (the symmetric part of) the Ricci tensor in
Weyl geometry (see appendix~\ref{appsec:Weyl} for a small introduction).
\par
Contracting the above integrability condition with $\gamma^{a}$ one finds that
\begin{equation}
  \label{eq:3}
  n\ \slashed{F}\epsilon \; =\; \mathrm{W}\ \epsilon \; ,
\end{equation}
which when combined with eq.~(\ref{eq:2}) leads to 
\begin{equation}
  \label{eq:4}
  \textstyle{1\over 2}\left[\
         \mathrm{W}_{(ab)} \; -\; \textstyle{1\over n}\mathrm{W}g_{ab}
  \ \right]\ \gamma^{b}\epsilon \; =\; 0\; .
\end{equation}
In the Riemannian setting the above is enough to conclude that if we find a
spinor $\epsilon$ satisfying eq.~(\ref{eq:1}), then the underlying geometry is
Einstein-Weyl. In the non-Riemannian setting this conclusion is not true:
experience from the classification of supersymmetric solutions to supergravity
theories shows instead that there are two quite different cases to be
considered, namely the timelike or the null case. The sexer of these two cases
is the norm of a particular vector-bilinear built out of the Killing spinor,
which can be shown to be either zero or positive, hence the naming of the
cases. The minimal set of equations of motion that need to be imposed in order
to guarantee that all EOMs are satisfied, is different in each case: in the
timelike case a supersymmetric field configuration automatically satisfies the
EW condition, whereas in the null case the minimal set consists of only one
component of the EW condition, namely the one lying in the double direction of
the null vector-bilinear.
\par
Seeing the similarity of the integrability condition of the spinorial rule
with the geometric constraint for EW spaces, it should not come as a surprise
that eq.~(\ref{eq:1}) is invariant under the following Weyl transformations
\begin{equation}
  \label{eq:16}
  \begin{array}{lclclcl}
    g & =& e^{2w}\tilde{g} &\hspace{.4cm},\hspace{.4cm}& e^{a} & =& e^{w}\tilde{e}^{a} \; ,\\
    A & =& \tilde{A}+dw & ,& \theta_{a} & =& e^{-w}\tilde{\theta}_{a} \; , \\
    \epsilon & =& e^{\alpha w}\tilde{\epsilon} & ,& \alpha & =& \textstyle{\frac{4-n}{4}} \; .
  \end{array}
\end{equation}
This Weyl symmetry can in fact be used to obtain the r.h.s.~of
eq.~(\ref{eq:1}), which would otherwise have to be wild-guessed: the Weyl
connection, eq.~(\ref{eq:W2}), in the spinorial representation is given by
$\mathtt{D}_a=\nabla_a-\frac{1}{2}\gamma_{ab}A^b$ 
, allowing us to rewrite eq.~(\ref{eq:1}) as
$\mathtt{D}_a\epsilon=\frac{4-n}{4}\,A_a\,\epsilon$. In other words, we are
dealing with a weighted Killing spinor in Weyl geometry.
\par
The next step in the analysis is to define the bilinear $\hat{L} =
L_{\mu}dx^{\mu} = \overline{\epsilon}\gamma_{\mu}\epsilon\ dx^{\mu}$, which
(as shown in appendix~\ref{appsec:spinors}) is a real 1-form and, for a Lorentzian
spacetime, is either timelike, $g(L,L)>0$ in our conventions, or null,
$g(L,L)=0$.  Independently of these details, however, we can always derive
from the spinorial equation (\ref{eq:1}) the following differential rule for
the bilinear
\begin{equation}
  \label{eq:33}
  \nabla_{a}L_{b} \; =\ \frac{4-n}{2}\ A_{a}L_{b} \ -\ L_{a}A_{b} \ +\ \imath_{L}A\ g_{ab}\; ,
\end{equation}
whose totally antisymmetric part reads
\begin{equation}
  \label{eq:6}
  d\hat{L} \; =\; \frac{6-n}{2}\ A\wedge \hat{L} \; ,
\end{equation}
singling out the $n=6$ case as special, as $\hat{L}$ is then closed.
\par
We shall start the analysis by considering the timelike case.
\section{Timelike solutions}
\label{sec:timelike}
Suppose that $L$ is timelike and define $f\equiv g(L,L)$. We can
straightforwardly use eq.~(\ref{eq:33}) to find
\begin{equation}
  \label{eq:34}
   df \; =\; (4-n)\ A\ f   \; ,
\end{equation}
so that, as long as $n\neq 4$, the Weyl structure is exact and any
supersymmetric EW-space is equivalent to a metrical space allowing for a
parallel spinor. Bryant \cite{art:Bryant} has classified all the
pseudo-Riemannian spaces admitting covariantly-constant spinors for a
different number of dimensions. Then, this prescribes the \emph{timelike}
Einstein-Weyl metrics with Lorentzian signature in dimensions three (flat),
five and six ($g=\mathfrak{R}^{1,n-5} \times \tilde{g}$, where $\tilde{g}$ is
a 4-dimensional Ricci-flat K\"ahler manifold). A general study for the
remaining dimensions is still an open problem, as far as we know. However,
Galaev \& Leistner \cite{art:galaev2008} provide a partial answer by giving a
blueprint for the geometry of simply-connected, complete Lorentzian spin
manifolds that admit a Killing spinor (see theorem 1.3 therein).
\par

For the $n=4$ case, we use the same building blocks as in
ref.~\cite{Meessen:2009ma} to set up the whole calculus of spinor
bilinears. We deal with the spinor structure of $N=2$ $n=4$ supersymmetry,
which allows us to decompose a Dirac spinor in $n=4$ as a sum of two Majorana
spinors, which we can project onto the anti-chiral part, denoted
$\epsilon_{I}$ ($I=1,2$), and the chiral part, denoted by $\epsilon^{I}$. Here
the position of the $I$-index indicates exclusively the chirality, and the
chiralities are interchanged by complex conjugation {\em i.e.\/}
$(\epsilon_{I})^{*}=\epsilon^{I}$, so the theory has two independent spinors.
Doing this decomposition, the rule eq.~(\ref{eq:1}) can be written as
\begin{equation}
  \label{eq:35}
  \nabla_{a}\epsilon_{I} \; =\; \textstyle{1\over 2}\gamma_{ab}A^{b}\epsilon_{I}
  \hspace{.5cm}\mbox{and}\hspace{.5cm}
  \nabla_{a}\epsilon^{I} \; =\; \textstyle{1\over 2}\gamma_{ab}A^{b}\epsilon^{I} \; .
\end{equation}
Using the spinors one can then construct (see ref.~\cite{Meessen:2009ma}) a
complex scalar $X\equiv\frac{1}{2}\varepsilon^{IJ}\bar{\epsilon}_I\epsilon_J$,
3 complex 2-forms $\Phi^{x}$ ($x=1,2,3$) that will not play any r\^{o}le in
what follows, and 4 real 1-forms $V^{a}=i\bar{\epsilon}^I \gamma^a
\epsilon_I$.  These 4 1-forms form a linearly independent base and can be used
to write the metric, $g$, as
\begin{equation}
  \label{eq:36}
  4|X|^{2}\ g \; =\; \eta_{ab}\ V^{a}\otimes V^{b} \; ,
\end{equation}
whence $V^{0}\sim L$. Given the definitions of the bilinears we can calculate
\begin{eqnarray}
  \label{eq:37}
  dX  & =& 0\; ,\\
  \label{eq:37a}
  dV^{a} & =& A\wedge V^{a} \; ,
\end{eqnarray}
meaning that $X$ is just a complex constant.  The integrability condition of
eq.~(\ref{eq:37a}) is $F\wedge V^{a} = 0$ which, due to the
linear-independency of the $V^{a}$ implies that $F=0$.
Locally, then, we can transform $A$ to zero and introduce coordinates $x^{a}$
such that $V^{a}= 4|X|^{2}\ dx^{a}$, resulting in a Minkowski metric. Whence,
in $n=4$ a \emph{timelike} supersymmetric Lorentzian EW space is locally
conformal to Minkowski space.
\\
\par
The conclusion then w.r.t.~the \emph{timelike} solutions to the rule
(\ref{eq:1}) is that they are trivial in the sense that they are always
related by a Weyl transformation to a Lorentzian space admitting Killing
spinors, {\em i.e.\/} spinors satisfying the rule $\nabla_{a}\epsilon =0$.
\par
The analysis of the null cases is more involved, mainly due to a lack of systematics
in the bilinears, the exception being the vector bilinear $L$ as 
one can see from eq.~(\ref{eq:33}), but also because the bilinear approach to classification
of supersymmetric solutions becomes unwieldy for $n>6$.
In stead of attempting to do a complete analysis in all the cases where the bilinear approach
can be applied, we shall analyse the cases $n=4$ and $n=6$ explicitly, and then
give some generic comments in section (\ref{sec:general}).
\section{Null $N=1$ $n=4$ solutions}
\label{sec:N1D4}
The natural starting point, seeing the explicit case treated in the foregoing section, would be 
the null case in $n=4$ $N=2$. Prior experience with this case in supergravity, however, shows that
this case is related to the simpler case of $n=4$ $N=1$ supergravity \cite{art:ortin2008},
a theory for which the vector bilinear $L$ is automatically a null vector.
In $n=4$ $N=1$ sugra the spinor is a Weyl spinor, and one can see that the KSE (\ref{eq:1})
is compatible with the truncation of $\epsilon$ to a chiral spinor, and in this section we shall
henceforth take $\epsilon$ to be a Weyl spinor.
\par
The first rule we can derive for the bilinear is
\begin{equation}
  \label{eq:5}
  \nabla_{a}L_{b} \; =\; - L_{a}A_{b} \ +\ \imath_{L}A\ g_{ab} \; ,
\end{equation}
which is already enough to see that $L^{\flat}$ is a geodesic null vector.
The antisymmetric and symmetric parts of the above equation read
\begin{eqnarray}
  \label{eq:7}
  d\hat{L} & =& A\wedge \hat{L} \; , \\
   \label{eq:8}
  \nabla_{(a} L_{b)} & =& - A_{(a}L_{b)}  \; +\; \textstyle{1\over 3}\ \nabla\cdot L\ g_{ab}\; .
\end{eqnarray}
\par
There is another bilinear that can be constructed \cite{art:ortin2008}, which
is a 2-form defined as $\Phi_{ab}=\overline{\epsilon}\gamma_{ab}\epsilon$ and
using the propagation rule we can deduce
\begin{equation}
  \label{eq:17}
  \nabla_{a}\Phi_{bc} \; =\; 2\Phi_{a[b}A_{c]} \ -\ 2 g_{a[b}\Phi_{c]d}A^{d} \; ,
\end{equation}
which through antisymmetrisation gives rise to
\begin{equation}
  \label{eq:18}
  d\Phi \; =\; 2A\wedge\Phi \; .
\end{equation}
\par
Eq.~(\ref{eq:7}) implies that $\hat{L}\wedge d\hat{L} =0$, whence $\hat{L}$ is
hypersurface orthogonal, and we can use the Frobenius theorem to introduce two
real functions $u$ and $P$ such that $\hat{L}=e^{P}du$. Since by
eq.~(\ref{eq:7}) above $\hat{L}$ has gauge charge 1 under $A$, we can perform
a Weyl-gauge transformation to take $P=0$, as to obtain $\hat{L}=du$. This
further implies that $A=\Upsilon\ \hat{L}$, where $\Upsilon$ is a real
function whose coordinate dependence needs to be deduced, and also
$\imath_{L}A=0$.  {}Furthermore, we see that $d^{\dagger} \hat{L}=0$ and
$\nabla_{L}L=0$, {\em i.e.\/} $L$ is the tangent vector to an affinely
parametrised null geodesic.
\par
Observe that we can apply the same reasoning for eq.~(\ref{eq:6}) in
dimensions different from six: as long as $n\neq 6$ we can always use a Weyl
transformation as to fix $\hat{L}=du$ and write $A=\Upsilon\ \hat{L}$. The
fact that in the case $n=6$ the 1-form $\hat{L}$ is automatically closed has
profound implications, as will be shown in section~(\ref{sec:D6chiral}).
\par
Having fixed the Weyl symmetry, we can introduce a normalised null tetrad
\cite{Penrose:1985jw} and a corresponding coordinate representation by
\begin{equation}
\label{eq:19}
\begin{array}{lclclcl}
\hat{L} 
& = & 
du &\hspace{.4cm},\hspace{.4cm}& L & =& \partial_{v} \; , \\
\hat{N} 
& = & 
dv + Hdu +\varpi dz + \bar{\varpi}d\bar{z} & ,& N & =& \partial_{u}\ -\ H\partial_{v} \; , \\
\hat{M} 
& = & 
Udz & ,& M & =& -\bar{U}^{-1}\left( \partial_{\bar{z}} \ -\ \bar{\varpi}\partial_{v}\right) \; ,\\
\hat{\overline{M}} 
& = & 
\bar{U}d\bar{z} & ,& \overline{M} & =& -U^{-1}\left( \partial_{z} \ -\ \varpi\partial_{v}\right) \; ,
  \end{array}
\end{equation}
for which the metric reads
\begin{eqnarray}
\label{eq:20}
g & = & 
\hat{L}\otimes \hat{N} \ +\ \hat{N}\otimes \hat{L} \ -\
\hat{M}\otimes\hat{\overline{M}} \ -\ \hat{\overline{M}}\otimes \hat{M}
\hspace{1cm}\longrightarrow
\nonumber \\
ds^{2}   & = & 
2du\left( dv + Hdu + \varpi dz + \bar{\varpi}d\bar{z}\right) \ -\ 2|U|^{2} dzd\bar{z} \; .
\end{eqnarray}
A straightforward calculation shows that the constraint (\ref{eq:8}) implies that 
\begin{equation}
  \label{eq:21}
  \Upsilon = -\partial_{v}H \;\;\; ,\;\;\; 
  \partial_{v}\varpi \ =\ 0 \;\;\; ,\;\;\;
  \partial_{v}\bar{\varpi} \ =\ 0 \;\;\; ,\;\;\;
  \partial_{v}|U|^{2} \ =\ 0 \; ,
\end{equation}
so that the only $v$-dependence resides in the function $H$, and we determined
the gauge field $A$ in terms of $H$.
\par
In $N=1$ $n=4$ one can see that $\Phi = \hat{L}\wedge\hat{\overline{M}}$ (see
{\em e.g.\/} \cite[eq.~(70)]{Meessen:2009ma}). Combining this with
eq.~(\ref{eq:18}) we see that
\begin{equation}
  \label{eq:22}
  0 \ =\ \hat{L}\wedge d\hat{\overline{M}} \ =\ 
d\bar{U}\wedge d\bar{z}\wedge du \hspace{1cm}\mbox{whence}\;\;
  \bar{U} \ =\ \bar{U}(u,\bar{z}) \; . 
\end{equation}
This result means that we can take $U=1$ by a suitable coordinate
transformation $Z=Z(u,z)$ such that $\partial_{z}Z = U$, which leaves the
chosen form of the metric invariant.
\par
In order to finish the analysis, let us investigate eq.~(\ref{eq:17}). As
$A\sim \hat{L}$ we have that $\imath_{A}\Phi \sim \imath_{L}\Phi = 0$ and we
find that $\nabla_{a}\Phi_{bc} = 2\Upsilon\ \Phi_{a[b}L_{c]}$.  Combining this
with $\Phi_{ab} = 2L_{[a}\overline{M}_{b]}$ we find that
\begin{equation}
  \label{eq:23}
  0 \; =\; L_{[b|}\nabla_{a}\overline{M}_{|c]} \; ,
\end{equation}
which can be evaluated on the chosen coordinate basis to give
\begin{equation}
  \label{eq:24}
  0\; =\; \partial_{\bar{z}}\varpi \ -\ \partial_{z}\bar{\varpi} 
\hspace{1cm}\mbox{which implies:}\;\;
  \varpi \ =\ \partial_{z}B \; ,\; \bar{\varpi}\ =\ \partial_{\bar{z}}B \; ,
\end{equation}
where $B$ is a real function. As is well-known, one can then get rid of
$\varpi$ altogether by a suitable shift of the coordinate $v \rightarrow v-B$.
\par
The end result of this analysis is that, given the fact that the spinor
$\epsilon$ is taken to be a Weyl spinor, any solution\footnote{By solution we
  refer to a geometry that arises from the existence of a spinor that fulfills
  eq.~(\ref{eq:1}).}~to the equation~(\ref{eq:1}) is related by a Weyl
transformation to
\begin{eqnarray}
\label{eq:10}
ds^{2}_{(4)} & = & 
2du\left( dv \ +\ Hdu\right) \; -\; 2dzd\bar{z} \; , \\
A & = & 
-\partial_{v}H\ du \; ,
\end{eqnarray}
Actually, this metric is a special case of a more-general metric, referred to
as a Kundt metric in the physics literature (see appendix~(\ref{appsec:Kundt})
for more information), a type of metric that appears naturally in the null case of 
not only supergravity \cite{Brannlund:2008zf} solutions, but also fake supergravity
solutions\change{, see {\em e.g.\/} refs.~\cite{Meessen:2009ma,Gutowski:2009vb} and \cite{Grover:2009ms}}.
\par
At this point we would like to recall what was mentioned in
section~(\ref{sec:maths}) above about pseudo-Riemannian signatures and certain
EOMs (the EW conditions in this case) not having to be explicitly
checked. Since we are trying to give a prescription for EW spaces, we
obviously need to satisfy eq.~(\ref{eq:W5}). An explicit calculation shows
that the integrability conditions (\ref{eq:4}) are automatically satisfied,
with the only non-trivial component being $\mathrm{W}(N,N)\
L_{c}\gamma^{c}\epsilon$. Adapting the Fierz identities to the null case
scenario, one obtains the constraint $L_{c}\gamma^{c}\epsilon =0$ (see {\em
  e.g.} eq.~(5.12) of ref.~\cite{Bellorin:2005}), satisfying this way the
integrability
condition
, and hence we see that we have a  solution to the KSE.
\par
However, we still need to ensure that the local geometry (\ref{eq:10}) indeed
solves all EW conditions (\ref{eq:W5}), and we must therefore impose by hand
that $\mathrm{W}(N,N)=0$. A small calculation shows that this implies that $H$
must satisfy the following differential equation
\begin{equation}
  \label{eq:11}
  \partial_{u}\partial_{v}H \ -\ H\partial_{v}^{2}H  \; =\; \partial\bar{\partial}H \; . 
\end{equation}
We can find a four-dimensional generalisation of the Weyl-scalar-flat EW
geometry obtained by Calderbank \& Dunajski in \cite{art:Calderbank01SFlat} by
using a function $H$ of the form
\begin{equation}
 \label{eq:12b}  
 H\ =\ v\partial F +v\bar{\partial} \bar{F}+\bar{z}\partial_u F +z \partial_u \bar{F} \qquad \text{where } F=F(u,z)\ .
\end{equation}
It gives rise to a non-trivial EW space as long as $\partial^2 F\neq 0$.
\section{Null $N=(1,0)$ $n=6$ solutions}
\label{sec:D6chiral}
As in the foregoing section we will consider the spinor $\epsilon$ to be chiral
which not only implies that the vector bilinear is null, but also
that we can use the results of  Gutowski {\em et al.\/} \cite{Gutowski:2003rg},
who classified the
supersymmetric solutions of ungauged chiral supergravity in 6 dimensions,
{\em i.e.\/} minimal $n=6$ $N=(1,0)$ supergravity.
This theory is in itself quite curious, and so are the
spinor bilinears: there is only a null vector $L$ and a triplet of selfdual
3-forms $\Phi^{r}_{(3)}$ ($r=1,2,3$). These bilinears are defined by
\begin{equation}
\label{eq:40}
\begin{array}{lclclcl}
L_{a} & \equiv& 
-\varepsilon^{IJ}\ \epsilon_{I}^{c}\gamma_{a}\epsilon_{J} &\;\;\; , \;\;\;& 
\epsilon_{I}^{c}\gamma_{a}\epsilon_{J} & =& 
-\textstyle{1\over 2}\ \varepsilon_{IJ}\ L_{a} \; ,\\
 && && && \\
\Phi^{r}_{abc} &\equiv& i\left[\sigma^{r}\right]^{IJ}\ 
\epsilon_{I}^{c}\gamma_{abc}\epsilon_{J} & ,&
\epsilon_{I}^{c}\gamma_{abc}\epsilon_{J} & =& 
\textstyle{i\over 2}\ \left[\sigma^{r}\right]_{IJ}\ \Phi^{r}_{abc} \; , 
  \end{array}
\end{equation}
where $\epsilon^{c}=\epsilon^{T}\mathcal{C}$ means the Majorana conjugate.

These bilinears satisfy the following Fierz-relations
\begin{eqnarray}
  \label{eq:39}
  L_{a}L^{a} & =& 0 \; , \\
  \label{eq:39a}
  \imath_{L}\Phi^{r}_{(3)} & =& 0 \hspace{.5cm}\longrightarrow\hspace{.5cm} \hat{L}\wedge\Phi^{r}_{(3)} \ =\ 0  \; , \\
  \label{eq:39b}
  \Phi^{r\ fab}\Phi^{s}_{fcd} & =& 4\delta^{rs}\ L^{[a}L_{[c}\ \eta^{b]}_{d]} 
            \ -\ \varepsilon^{rst}L^{[a|}\Phi^{t\ |b]}{}_{cd}
            \ +\ \varepsilon^{rst}L_{[c}\Phi^{t\ ab}{}_{d]} \; .
\end{eqnarray}
Seeing eqs.~(\ref{eq:39a}) and (\ref{eq:39b}) we find that $\Phi^{r}_{(3)} =
\hat{L}\wedge\mathsf{K}^{r}_{(2)}$ with $\imath_{L}\mathsf{K}_{(2)}^{r}=0$.
\par
Using the definitions of the bilinears we can use the rule eq.~(\ref{eq:1}) to
calculate the effect of parallel-transporting them. The results is that for an
arbitrary vector field $X$ we have
\begin{eqnarray}
\label{eq:43}
\nabla_{X}\hat{L} & = & 
-\imath_{X}A\ \hat{L} \ -\ \imath_{X}\hat{L}\ A \ +\ \imath_{L}A\ \hat{X} \; ,\\
\label{eq:43a}
\nabla_{X}\Phi^{r} &  = & 
-\imath_{X}A\ \Phi^{r} \ +\ \hat{X}\wedge\imath_{A^{\flat}}\Phi^{r} \ -\ A\wedge\imath_{X}\Phi^{r} \; ,
\end{eqnarray}
{}From eq.~(\ref{eq:43}) it is clear that $L$ is a null geodesic, 
{\em i.e.\/} $\nabla_{L}L=0$, and, as we already knew from (\ref{eq:6}), $d\hat{L}=0$.
\par
At this point then we can, as before, introduce a Vielbein adapted to the null
nature of $L$ in terms of the natural coordinates $v$, $u$ and $y^{m}$
($m=1,\ldots ,4$) as
\begin{equation}
\label{eq:KundtCurv1a}
\begin{array}{lclclcl}
E^{+} 
& = & 
du &\hspace{.4cm},\hspace{.4cm}& \theta_{+} & =& \partial_{u} \ -\
H \partial_{v} \; , 
\\
E^{-} 
& = & 
dv + Hdu + S_{m}dy^{m} & ,& \theta_{-} & =& \partial_{v} \; ,
\\
E^{i} 
& = & {e_{m}}^{i}\ dy^{m} & ,& \theta_{i} & =& {e_{i}}^{m}\left[ \partial_{m} \ -\ S_{m}\partial_{v}\right] \; ,
  \end{array}
\end{equation}
where $\hat{L}\equiv E^{+}$ and $L\equiv \theta_{-}$.  As usual we can then
define the metric on the base space by $\mathsf{h}_{mn}\equiv
{e_{m}}^{i}{e_{n}}^{i}$ and we can write the full 6-dimensional Kundt metric
as
\begin{equation}
  \label{eq:44}
  ds_{(6)}^{2} \; =\; 2du\left( dv\ +\ Hdu\ +\ \hat{S}\right) \ -\ \mathsf{h}_{mn}\,dy^{m}dy^{n} \; .
\end{equation}
\par
We can expand the 2-forms as $2\ \mathsf{K}^{r}\equiv \mathsf{K}^{r}_{ij}
E^{i}\wedge E^{j}$ w.r.t.~the above Vielbein, and by choosing the light-cone
directions such that $\varepsilon^{+-1234}=1=\varepsilon^{1234}$, we see that
$\star_{(4)}\mathsf{K}^{r} = -\mathsf{K}^{r}$.  Defining the $(1,1)$-tensors
$\mathsf{J}^{r}$ by means of $\mathsf{h}(\mathsf{J}^{r}X,Y)\equiv
\mathsf{K}^{r}(X,Y)$, we can see that eq.~(\ref{eq:39b}) implies
\begin{equation}
  \label{eq:41}
  \mathsf{J}^{r}\mathsf{J}^{s} \; =\; -\delta^{rs} \, +\, \varepsilon^{rst}\ \mathsf{J}^{t} \; ,
\end{equation}
so that the 4-dimensional base space is always going to be an {\em almost
  quaternionic manifold}.
\par
At this point we will fix part of the Weyl gauge symmetry by imposing the
gauge-fixing condition $\imath_{L}A=0$ and consequently we can expand the
gauge field as
\begin{equation}
  \label{eq:45}
  A \; =\; \Upsilon\ \hat{L} \; +\; \mathsf{A}_{m}dy^{m} \; .
\end{equation}
Using this expansion and the explicit form of the Vielbein in terms of the
coordinates, we can analyse eq.~(\ref{eq:43}), resulting in
\begin{eqnarray}
  \label{eq:46}
  \Upsilon & =& -\textstyle{1\over 2}\ \partial_{v}H \; , \\
  \label{eq:46a}
  \partial_{v}\hat{S} & =& -2\ \mathsf{A}\; ,\\
  \label{eq:46b}
  0 & =& \partial_{v}\mathsf{h}_{mn} \; .
\end{eqnarray}
Contrary to what is usually the case in (fake) supergravities, we do not know
the full $v$-dependence of $H$ and therefore we cannot completely fix the
$v$-dependence of the unknowns. The above results comprise all the information
contained in eq.~(\ref{eq:43}).
\\
\par
In order to analyse the content of eq.~(\ref{eq:43a}) we first take $X=L$ to
find that $\nabla_{L}\Phi^{r}=0$, which when evaluated in the chosen
coordinate system implies $\partial_{v}\mathsf{K}^{r}_{mn}=0$. This innocuous
result fixes, however, the $v$-dependence of $\mathsf{A}$: from the totally
antisymmetric part of eq.~(\ref{eq:43a}) one obtains
\begin{equation}
  \label{eq:47}
  d\Phi^{r}\; =\; 2A\wedge\Phi^{r} \;\longrightarrow\;
  0\ =\ \hat{L}\wedge\left[ \mathsf{d}\mathsf{K}^{r}\ -\ 
2\mathsf{A}\wedge\mathsf{K}^{r}\ \right] \; ,
\end{equation}
where we introduced the exterior derivative on the base space
$\mathsf{d}\equiv dy^{m}\partial_{m}$.  As the $\mathrm{K}$s are
$v$-independent and $\hat{L}=du$, we see that the consistency of the above
equation requires $\mathsf{A}$ to be $v$-independent. Then, we also obtain
from eq.~(\ref{eq:46a}) that
\begin{equation}
  \label{eq:48}
  \hat{S} \; =\; -2v\ \mathsf{A} \, +\, \varpi \hspace{2cm}(\partial_{v}\varpi_{m}=0 )\; .
\end{equation}
It should be clear from eq.~(\ref{eq:47}) that the $y$-dependence of the
$\mathrm{K}$s is given by the equation
\begin{equation}
\label{eq:49}
\mathsf{d}\mathsf{K}^{r}\ =\ 2\mathsf{A}\wedge\mathsf{K}^{r}
\hspace{1cm}\mbox{whose integrability condition reads}\hspace{1cm}
\mathsf{F}\wedge\mathsf{K}^{r} \ =\ 0\; ,
\end{equation}
where we defined $\mathsf{F}=\mathsf{d}\mathsf{A}$. Actually, the last equation 
implies, as one can easily verify, that $\mathsf{F}$ is selfdual, {\em i.e.\/}
$\star_{(4)}\mathsf{F}=\mathsf{F}$, whence $\mathsf{A}$ is a selfdual
connection or in physics-speak an $\mathbb{R}$-instanton.
\par
The analysis of eq.~(\ref{eq:43a}) in the direction $X=\theta_{+}$ is
straightforward and leads to the following constraints on the spin connection
\begin{eqnarray}
\label{eq:51}
\omega_{+-k}\ \mathsf{K}^{r}_{kj} 
& = & 
-\mathsf{A}_{k}\ \mathsf{K}^{r}_{kj} \; ,
\\
\label{eq:51a}
0 & = & 
\omega_{+i}{}^{k}\ \mathrm{K}^{r}_{kj} \; +\; \omega_{+j}{}^{k}\ \mathrm{K}^{r}_{ik} \; .
\end{eqnarray}
By using the results in appendix~\ref{appsec:Kundt}, we see that
eq.~(\ref{eq:51}) is automatically satisfied. A small investigation in
eq.~(\ref{eq:51a}) shows that it implies the base space 2-form $\omega_{+ij}\
E^{i}\wedge E^{j}$ to be selfdual! Coupling this observation with
eq.~(\ref{eq:KundtCurv2d}) and taking into account $\mathsf{F}$'s selfduality,
we see that the base space 2-form $2\Omega = \Omega_{ij}\ E^{i}\wedge E^{j}$,
whose components are defined by
\begin{equation}
\label{eq:52}
\Omega_{ij}\; \equiv\; 2\mathsf{D}_{[i}\varpi_{j]} \ +\ 2e_{[i}{}^{m}\partial_{u}e_{j]m}
\hspace{2cm}(\mbox{where:}\;\; 
\mathsf{D}\varpi \equiv \mathsf{d}\varpi\ -\ 2\mathsf{A}\wedge\varpi )\; ,
\end{equation}
has to be selfdual, {\em i.e.\/} $\star_{(4)}\Omega = \Omega$.
\par
In order to completely drain eq.~(\ref{eq:43a}) of information we need to
consider $X$ lying on the base space. Let $\mathsf{X}$ be such a vector. Then,
we find that
\begin{equation}
\label{eq:50}
\nabla^{(\lambda )}_{\mathsf{X}}\ \mathrm{K}^{r} \; =\; 
\mathsf{X}^{\sharp}\wedge\ \star_{(4)}\left[\ \mathsf{A}\wedge\mathsf{K}^{r}\ \right]
 \ -\ \mathsf{A}\wedge \imath_{\mathsf{X}}\mathsf{K}^{r} \; ,
\end{equation}
where $\nabla^{(\lambda )}$ is the ordinary spin connection on the base space
using the $\lambda$s in eq.~(\ref{eq:KundtCurv2d}).  {}Following
ref.~\cite{Grover:2008jr} we can then introduce a torsionful connection
$\overline{\nabla}_{\mathsf{X}}\mathsf{Y}\equiv \nabla^{(\lambda
  )}_{\mathsf{X}}\mathsf{Y}\ -\ \mathsf{S} _{\mathsf{X}}\mathsf{Y}$ with the
torsion being totally antisymmetric and proportional to the Hodge dual of the
$\mathbb{R}$-gauge field, {\em i.e.\/}
\begin{equation}
\label{eq:53}
\mathsf{h}\left( \mathsf{S}_{\mathsf{X}}\mathsf{Y},\mathsf{Z}\right) \;
\equiv\; 
-\left[\star_{(4)}\mathsf{A}\right]\ \left( \mathsf{X},\mathsf{Y},\mathsf{Z}\right) \; ,
\end{equation}
such that eq.~(\ref{eq:50}) can be written compactly as
$\overline{\nabla}\mathsf{K}^{r}=0$. Almost quaternionic manifolds admitting a
torsionful connection parallelising the almost quaternionic structure are
called {\em Hyper-K\"ahler Torsion manifolds}, HKT manifolds for short, a name
that first appeared in \cite{Howe:1996} to describe the geometry of
supersymmetric sigma-model manifolds with torsion \cite{Gates:1984}.
\par
As pointed out in ref.~\cite{Grover:2008jr}, we can make use of the residual
Weyl symmetry in eq.~(\ref{eq:16}) with $w=w(y)$, {\em i.e.\/} a Weyl
transformation depending only on the coordinates of the base space, to
gauge-fix the condition $\mathsf{d}^{\dagger}\mathsf{A}=0$.  This immediately
implies that the torsion $\mathsf{S}$ is closed, and the resulting
mathematical 4-dimensional structure is called a {\em closed HKT
  manifold}. Let us mention, even though it will not be needed, that the
coordinate transformation $v\rightarrow v+\Lambda (y)$, induces the 'gauge'
transformation $\varpi\rightarrow \varpi + \mathsf{D}\Lambda$.
\\
\par
Thus far, the analysis has shown that the pair $(g,A)$ admits a solution to
eq.~(\ref{eq:1}) iff $g$ is the metric of a Kundt wave whose base space is a
$u$-dependent family of HKT-spaces. Given such a family of HKT spaces we can
find the 1-form $\varpi$ by imposing selfduality of the 2-form $\Omega$ in
eq.~(\ref{eq:52}) and then the only indeterminate element of the metric is the
wave profile $H$. This analysis has given us the necessary conditions for the
existence of a non-null spinor satisfying eq.~(\ref{eq:1}).  It remains to be
checked that they are also sufficient by direct substitution into
eq.~(\ref{eq:1}).
\par
A quick calculation of the $(-)$ component, leads to $\theta_{-}\epsilon =0$,
whence the spinor is $v$-independent.  The $(+)$-component leads, after using
the constraint $\gamma^{+}\epsilon =0$, to
\begin{equation}
\label{eq:25}
\partial_{u}\epsilon \; =\; -\textstyle{1\over 4}\ T_{ij}\ \gamma^{ij}\epsilon \; =\; 0 \; ,
\end{equation}
where the last step follows from the selfduality of $T$ (see eq.~(\ref{eq:9}))
and the chirality of the spinor $\epsilon$. We conclude that the spinor is
also $u$-independent.  Giving the $i$ components of eq.~(\ref{eq:1}) a similar
treatment we end up with
\begin{equation}
  \label{eq:55}
  \nabla_{i}^{(\lambda )}\epsilon \; =\; \textstyle{1\over 2}\ \tilde{\gamma}_{ij}\ \mathsf{A}_{j}\epsilon \; ,
\end{equation}
where we have defined $\tilde{\gamma}^{i}\equiv i\gamma^{i}$, so
$\{\tilde{\gamma}^{i},\tilde{\gamma}^{j}\} = 2\delta^{ij}$, in order to obtain
a purely Riemannian spinorial equation.
\par
As one  can readily see from  eq.~(\ref{eq:1}), the above  equation is nothing
more than  its Riemannian  version for four-dimensional  spaces: this  kind of
spinorial equations  was studied by  Moroianu in ref.~\cite{art:moroianu1996a}
who investigated  Riemannian Weyl geometries admitting  spinor fields parallel
w.r.t.~the  Weyl connection.   {}For $n\neq  4$ he  found that  any  such Weyl
structure was  closed, whereas  in $n=4$ he  found the  HKT structure outlined
above. Furthermore, he showed that, if the 4-dimensional space is compact, then
the HKT structure is conformally related to either a flat torus, a K3 manifold
or the Hopf surface $S^{1}\times S^{3}$ with the standard, locally flat metric
(see {\em e.g.\/} \cite{Gibbons:1997iy}).
\par
The integrability condition of eq.~(\ref{eq:55}) implies that the Ricci tensor
of the metric $\mathsf{h}$ has to satisfy
\begin{equation}
  \label{eq:58}
  \mathsf{R}(\mathsf{h})_{ij} \; =\; 2\nabla^{(\lambda )}_{(i}\mathsf{A}_{j)} \ +\ 2\mathsf{A}_{i}\mathsf{A}_{j}
              \ +\ \mathsf{h}_{ij}\left(\ \nabla^{(\lambda )}_{i}\mathsf{A}_{i}\ -\ 2\mathsf{A}^{2}\ \right) \; ,
\end{equation}
which, by comparison with eq.~(\ref{eq:14}), is equivalent to saying that the pair $(\mathsf{h},\mathsf{A})$ forms a Ricci-flat Weyl geometry {\em i.e.\/} $\mathrm{W}_{(ij)}=0$.\\
\par
As we did in section~(\ref{sec:N1D4}), we impose the Einstein-Weyl equations in
those directions in which it is not trivially satisfied, {\em i.e.} in the
($++$)-direction. This, in turn, fixes the function $H$, which was otherwise
unknown. At this point, however, we would like to impose the simplifying
restriction that the HKT structure on the base space does not depend on
$u$. The motivation for this simplifying adjustment has to do with the
difficulty of finding analytic solutions to the differential equation
resulting from a $u$-dependent base space. A calculation of the
$(++)$-components of the E-W equations then shows that
\begin{equation}
  \label{eq:59}
  2\theta_{+}\theta_{-}H \ +\ \left(\theta_{-}H\right)^{2} \; =\; 
      \left(\ \nabla^{(\lambda )}_{i} - S_{i}\theta_{-} - 4\mathsf{A}_{i}\ \right)
      \left(\ \partial_{i} - S_{i}\theta_{-} -2\mathsf{A}_{i}\ \right)\ H \; ,
\end{equation}
where we have allow for a $u$-dependence of $H$.
\par
To summarise, any solution to the $N=(1,0)$ $n=6$ null scenario is once again
prescribed by a Kundt wave of the form eq.~(\ref{eq:44}) constrained by
eqs.~(\ref{eq:48}), (\ref{eq:52}) and (\ref{eq:59}), whose 4-dimensional base
space is given by a $v$-independent, closed HKT manifold subject to
eqs.~(\ref{eq:58}), and the gauge connection being that of an
$\mathbb{R}$-instanton.

\section{Remaining null cases}
\label{sec:general}
Having treated the null cases in $n=4$ and $n=6$, we are ready to make some
general comments on the null case in other dimensions. First of all, as was
pointed out in section~(\ref{sec:N1D4}), as long as $n\neq 6$ we can use a Weyl
transformation to introduce a coordinate $u$ such that $\hat{L}=du$ and then
also $A=\Upsilon \hat{L}$. Choosing the coordinate $v$ to be aligned with the
flow of $L\ (=\partial_{v})$, we can introduce the base space coordinates
$y^{m}$ ($m=1,\ldots , n-2$) and a Vielbein similar to the one in
eq.~(\ref{eq:KundtCurv1}), so that the metric is always of the form
\begin{equation}
  \label{eq:15}
  ds^{2}_{(n)} \; =\; 2du\left( dv \ +\ Hdu\ +\ S_{m}dy^{m}\right) \; -\; 
\mathsf{h}_{mn}\,dy^{m}dy^{n}\ ,
\end{equation}
where $\mathsf{h}_{mn}\equiv e^i_m e^i_n$. This is again a Kundt metric, and
evaluating the symmetric part of eq.~(\ref{eq:33}) in this coordinate system,
we get the following restrictions
\begin{equation}
  \label{eq:26}
  \Upsilon \ =\ -\frac{2}{n-2}\ \partial_{v}H 
    \;\; ,\;\; \partial_{v}S_{m} \ =\ 0 
    \;\; ,\;\; \partial_{v}\mathsf{h}_{mn} \ =\ 0 \; ,
\end{equation}
so that the whole $v$-dependence resides in $H$ and $\Upsilon$ only.
Following the convention in section~(\ref{sec:D6chiral}), we shall call the
$v$-independent part of $\hat{S}$ by $\varpi$, so that in the $n\neq 6$ case
we have $\hat{S}=\varpi$.
\par
With this information, and the constraint of $u$-independence imposed, we can
proceed to analyse the spinorial rule. The KSE in the $v$-direction is
automatically satisfied ({\em i.e.} $\partial_v \epsilon=0$) and the remaining
directions are
\begin{eqnarray}
  \label{eq:56}
  0 & =& \nabla_{i}^{(\lambda )}\ \epsilon \; , \\
  \label{eq:56a}
  \partial_{u}\epsilon & =& \textstyle{1\over 8}\ \left[ \mathsf{d}\varpi\right]_{ij}\gamma^{ij}\epsilon \; .
\end{eqnarray}
Eq.~(\ref{eq:56}) clearly states that the base space must be a Riemannian
manifold of special holonomy.  The integrability condition of the above two
equations then is that
\begin{equation}
  \label{eq:54}
  0\; =\; \left[ \nabla^{(\lambda )}_{i} (\mathsf{d}\varpi )_{kl}\right]\ \gamma^{kl}\epsilon 
 \hspace{.4cm}\mbox{which implies}\hspace{.4cm} 
 \left[ \mathsf{d}^{\dagger}\mathsf{d}\varpi\right]_{i}\gamma^{i}\epsilon \ =\ 0 \; ,
\end{equation}
so that $\mathsf{d}^{\dagger}\mathsf{d}\varpi =0$.\footnote{The same
  constraint can be obtained through explicit evaluation of the Einstein-Weyl
  equations.} Using the coordinate transformation $v\rightarrow v +\Lambda
(y)$ we can always take $\mathsf{d}^{\dagger}\varpi =0$, whence $\varpi\in
\mathrm{Harm}^{1}(\mathcal{B})$, {\em i.e.} $\varpi$ is a harmonic 1-form on
the base space.\footnote{Bochner's theorem states that any harmonic 1-form on a
  compact, oriented Ricci-flat manifold is parallel, which implies that in
  that case the Killing spinor is $u$-independent. In the non-compact case,
  however, there is no such theorem as can be envisaged by taking the base
  space to be $\mathbb{R}^{n-2}$ and to take $2\varpi \equiv
  f_{mn}x^{m}dx^{n}$, where the $f_{nm}$'s are constants.}
\par
Given this input, the condition for such a pair $(g,A)$ to be an Einstein-Weyl
manifold is
\begin{equation}
\label{eq:57}
2\partial_{u}\partial_{v}H \ -\ 
2H\partial_{v}^{2}H \ +\ 
2\textstyle{n-4\over n-2}\ \left(\partial_{v}H\right)^{2} \; =\;
  -\left( \nabla^{(\lambda )} - \varpi\right)^{i}\ \theta_{i}H  \; .
\end{equation}
The factor on the r.h.s.~of the above equations becomes, in the $\varpi =0$
limit, the d'Alembertian on the base space, and we make contact with
eq.~(\ref{eq:11}). This shows that the $n=4$ case is a subcase of the general
one studied in this section, where one was allowed to use the 2-form $\Phi$ to
get rid of $\hat{S}$. $n=6$, however, is an independent case where the
characteristic behaviour of the theory in that dimension (see {\em e.g.}
eq.~(\ref{eq:6})) nurtures the HKT structure.
\section{Summary and conclusions}
\label{sec:conclusions}
In this work we have presented a characterisation of supersymmetric
Einstein-Weyl spaces with Lorentzian signature in $n$ arbitrary dimensions. We
have done this by making use of the techniques developed for the
classification of supergravity solutions. In particular, we assumed the
existence of a spinor $\epsilon$ satisfying eq.~(\ref{eq:1}). It is in this
sense that our solutions have a supersymmetric character. We then proceeded to
build and analyse the bilinears that can be constructed from $\epsilon$, which
shape the resulting geometry.
\par
We have found that (for most dimensions) those spaces arising from a vector
bilinear which is timelike are trivial, in the sense that they are conformally
related to a space admitting a Killing spinor. The odd duck in the pond is the
4-dimensional case, for which the only \emph{timelike} solution actually turns
out to be Minkowski space, which coincides with which was already know for
parallel spinors \cite{art:Bryant}. The null case solutions are given by a
Kundt metric and a prescribed Weyl gauge field. It is worth mentioning that
the special structure of the $n=6$ case determines that the base space is
given by a closed {\em Hyper-K\"ahler Torsion} manifold. 
\par
As a closing paragraph let us consider the case $n=3$: in that case one can see that eq.~(\ref{eq:57}),
once one takes into account the fact that one perform coordinate transformations such that $\mathsf{h}=1$ 
and $\varpi =0$, corresponds to the dispersionless Kadomtsev-Petviashvili equation.
As shown in ref.~\cite[sec.~10.3.1.3]{boek:Dunajski}, the thus obtained class of 3-dimensional EW spaces is the unique class of 3-dimensional
EW spaces of Lorentzian signature admitting a weighted covariantly constant null vector.
{}Furthermore,
the supersymmetric class can be obtained by the Jones-Tod construction on a conformal space of neutral signature
admitting an anti-selfdual Null-K\"ahler structure \cite{boek:Dunajski}, a geometric structure which admits a parallel spinor.
Evidently, there are n-dimensional EW spaces, as there are 3-dimensional examples, that are not supersymmetric, and it would be interesting
to get a better handle on them.
\section*{Acknowledgments}
This work has been supported in part by the Spanish Ministry of Science and
Education grant FPA2009-07692, a C.S.I.C.~scholarship JAEPre-07-00176, a
Ram\'on y Cajal fellowship RYC-2009-05014, the Princip\'au d'Asturies grant
IB09- 069, the Comunidad de Madrid grant HEPHACOS S2009ESP-1473, and the
Spanish Consolider-Ingenio 2010 program CPAN CSD2007-00042. PM wishes to thank
J.~de Medinaceli and L.~Fern\'andez Seivane for useful discussions and TO
wishes to thank M.M.~Fern\'andez for her unfaltering support. AP would like to
dedicate this work to the memory of his grandfather, Ram\'on Lozano Cid, who
passed away while this article was being completed, and whose great sense of
responsability and care for the family -even on his ultimate moments- is
something AP would like to live up to.
{\appendix
\section{A short introduction to Einstein-Weyl geometry}
\label{appsec:Weyl}
A Weyl manifold is a manifold $\mathcal{M}$ of dimension $n$, a conformal
class $[g]$ of metrics on $\mathcal{M}$, and a torsionless connection
$\mathtt{D}$, which preserves the conformal class, {\em i.e.\/}
\begin{equation}
  \label{eq:W1}
     \mathtt{D}\ g \; =\; 2 A \otimes g \; , 
\end{equation}
for a chosen reference $g\in [g]$ and $A \in \Omega(\mathcal{M})$.  Using the
above definition, we can express the connection $\mathtt{D}_{X}Y$ as
\begin{equation}
  \label{eq:W2}
      \mathtt{D}_{\mu}Y_{\nu} \; =\; \nabla^{g}_{\mu}Y_{\nu} \ +\
      \gamma_{\mu\nu}{}^{\rho}\ Y_{\rho} \hspace{.3cm}\mbox{with}\hspace{.3cm}
      \gamma_{\mu\nu}{}^{\rho} \ =\ 
                g_{\mu}{}^{\rho}A_{\nu} \ +\ g_{\nu}{}^{\rho}A_{\mu} \ -\ g_{\mu\nu}A^{\rho}
            \; ,
\end{equation}
where $\nabla^{g}$ is the Levi-Civit\`a connection for the chosen $g\in
[g]$. We define the curvature of this connection as usual, {\em i.e.\/}
$\left[ \mathtt{D}_{\mu},\mathtt{D}_{\nu}\right] Y_{\rho} =
-\mathtt{W}_{\mu\nu\rho}{}^{\sigma}Y_{\sigma}$, and which we can use to define
the associated Ricci curvature as $\mathtt{W}_{\mu\nu} \equiv
\mathtt{W}_{\mu\sigma\nu}{}^{\sigma}$.  A calculation shows that the Ricci
tensor is not symmetric, which was to be suspected as we have a connection
with non-vanishing contorsion, and we have
\begin{eqnarray}
  \label{eq:W3a}
     \mathtt{W}_{[\mu\nu]} & =& -\textstyle{n\over 2}\
     F_{\mu\nu}\hspace{2.5cm}\mbox{other notations:}\, F=dA
     =\rho^{\mathtt{D}} \; , \\
      & & \nonumber \\
     \label{eq:W3b}
     \mathtt{W}_{(\mu\nu)} & =& \mathtt{R}(g)_{\mu\nu}
           \ -\ (n-2) \nabla_{(\mu}A_{\nu)}
           \ -\ (n-2)\ A_{\mu}A_{\nu}
           \ -\ g_{\mu\nu}\left[
                        \nabla_{\sigma}A^{\sigma} 
                    \ -\ (n-2)\ A_{\sigma}A^{\sigma}
                   \right] \; .
\end{eqnarray}
The Ricci-scalar is then of course defined as $\mathtt{W}\equiv
\mathtt{W}_{\sigma}{}^{\sigma}$, which explicitly reads
\begin{equation}
  \label{eq:W4}
      \mathtt{W} \; =\; \mathtt{R}(g) \ -\ 2(n-1)\ \nabla_{\sigma}A^{\sigma}
               \, +\, (n-1)(n-2)\
               A_{\sigma}A^{\sigma} \; .
\end{equation}
The 1-form $A$ acts as gauge field gauging an $\mathbb{R}$-symmetry, and this
is also the reason why we have been talking about a conformal class of metrics
on $\mathcal{M}$. In fact, under a transformation $g_{\mu\nu}\rightarrow
e^{2w}\ g_{\mu\nu}$, we have that $A\rightarrow A +dw$ and
$\mathtt{W}\rightarrow e^{-2w}\mathtt{W}$, whereas
$\mathtt{W}_{\mu\nu\rho}{}^{\sigma}$ and $\mathtt{W}_{\mu\nu}$ are conformally
invariant. In this sense, we say that an EW structure is trivial if the field
strength $F=dA=0$, {\em i.e.} locally the Weyl connection is conformally
vanishing.
\\

\par
A Weyl manifold is said to be Einstein-Weyl if the curvatures satisfy
\begin{equation}
  \label{eq:W5}
  \mathtt{W}_{(\mu\nu)} \; =\; \frac{1}{n}\ g_{\mu\nu}\ \mathtt{W} \; .
\end{equation}
\par
A metric $g$ in the conformal class is said to be {\em standard} or {\em
  Gauduchon} if it is such that
\begin{equation}
  \label{eq:W4a}
  d\star A \ =\ 0 \hspace{.4cm}\mbox{or equivalently}\hspace{.4cm}
  \nabla_{\sigma} A^{\sigma}\ =\ 0\; ,
\end{equation}
where the $\star$ is taken w.r.t.~the chosen metric.  Gauduchon
\cite{art:gauduchon1984a} showed that on a compact EW manifold there always
exists a standard metric, and Tod \cite{art:Tod1992} then went on to show that
on compact EW manifolds this implies that $A^{\flat}$ is a Killing vector of
the metric, {\em i.e.\/} it generates an isometry of $g$.

\section{Spinors in $\mathrm{SO}(1,d-1)$}
\label{appsec:spinors}
On $\mathbb{R}^{1,n-1}$ we shall put the mostly negative metric $\eta
=\mathrm{diag}(+,[-]^{n-1})$ and take the $\gamma$-matrices to satisfy
\begin{equation}
  \label{eq:27}
  \left\{\ \gamma_{a},\gamma_{b}\ \right\} \; =\; 2\eta_{ab} \; .
\end{equation}
We use a unitary representation of the $\gamma$-matrices, which implies that
$\gamma_{0}^{\dagger} =\gamma_{0}$ and $\gamma_{i}^{\dagger} =-\gamma_{i}$. Choosing the 
Dirac conjugation matrix $\mathcal{D}=\gamma_{0}$, we define the Dirac conjugate of a
spinor $\psi$ by $\overline{\psi}\equiv \psi^{\dagger}\mathcal{D}$ and find that 
\begin{equation}
  \label{eq:28}
  \mathcal{D}\gamma_{a}\mathcal{D}^{-1} \; =\; \gamma_{a}^{\dagger}
  \hspace{.5cm}\mbox{and}\hspace{.5cm}
  \mathcal{D}\gamma_{ab}\mathcal{D}^{-1} \; =\; -\gamma_{ab}^{\dagger}
\end{equation}
\par
Defining the 1-form $L=L_{a}\ e^{a}$ by means of $L_{a}\equiv
\overline{\psi}\gamma_{a}\psi$ which is then automatically real:
\begin{equation}
  \label{eq:29}
  L_{a}^{*} \ =\ \overline{\ \overline{\psi}\gamma_{a}\psi\ }
               \ =\ \psi^{T} \left(\mathcal{D}\gamma_{a}\right)^{*}\ \psi^{*}
              \ =\ \psi^{\dagger} \left(\mathcal{D}\gamma_{a}\right)^{\dagger}\ \psi
              \ =\ \overline{\psi}\mathcal{D}^{-1}\gamma_{a}^{\dagger}\mathcal{D}^{\dagger}\psi
              \ =\ \overline{\psi}\gamma_{a}\psi \ =\ L_{a} \; ,
\end{equation}
where a perhaps expected $-1$ sign in the third step is absent as we are
dealing with classical (commuting) spinors.
\par
In terms of the components we have that
$L_{a}=\epsilon^{\dagger}\mathcal{D}\gamma_{a}\epsilon$ and it is clear that
$L_{0}=\epsilon^{\dagger}\epsilon$. Furthermore, we can always rotate the
spatial components of $L$ in such a way that only the first component is
non-vanishing. This then implies that
\begin{equation}
  \label{eq:31}
  g(L,L) \ =\ L_{0}^{2}\ -\ L_{1}^{2} \; .
\end{equation}
$L_{1}=\epsilon^{\dagger}\gamma_{01}\epsilon$ and if we combine this with
$\gamma_{01}^{\dagger}=\gamma_{01}$, $\gamma_{01}^{2}=1$ and
$\mathrm{Tr}(\gamma_{01} )=0$ we can use a $\mathrm{SO}(\lfloor n/2\rfloor )$
rotation to write $\gamma_{01}= \mathrm{diag}([+]^{ \lfloor n/2\rfloor } ,
[-]^{\lfloor n/2\rfloor })$. Decomposing the spinor w.r.t.~the structure of
$\gamma_{01}$ as $\epsilon^{t}=(v,w)$, where $v$ and $w$ are vectors in
$\mathbb{C}^{\lfloor n/2\rfloor }$, we see that
\begin{equation}
  \label{eq:32}
  L_{0} \ =\ |v|^{2}+|w|^{2} \; ,\;
  L_{1} \ =\ |v|^{2}-|w|^{2} \;\;\longrightarrow\;\;
  g(L,L) \ =\ 4|v|^{2}|w|^{2} \; ,
\end{equation}
which implies the positive-definiteness of $|L|^2$.\\
\par
In the derivation of the spinorial rule eq.~(\ref{eq:1}) we have not made any
particular assumption about the nature of the spinor $\epsilon$ which has been
taken to be a (general) plain Dirac spinor. In the construction of the
bilinears, however, it is wise to impose a bit more structure on $\epsilon$;
this naturally leads one to investigate the compatibility of eq.~(\ref{eq:1})
with the conditions for the existence of a Weyl, Majorana, Majorana-Weyl
{\em {\&}c.\/}~spinor, a question that is answered affirmatively.
\section{Kundt metrics}
\label{appsec:Kundt}
\change{
A Kundt metric is a type of wave-like metric that allows for an expansion, shear and twist-free geodesic null-vector
\cite{Coley:2005sq} and were first studied in the 
arbitrary-$n$ case in refs.~\cite{Podolsky:2008ec} and \cite{art:juist}.
The line-element can always be taken to be
}
\begin{equation}
  \label{eq:KundtMetric}
ds^2=\hat{E}^+ \otimes \hat{E}^-+\hat{E}^- \otimes \hat{E}^+ - \hat{E}^x \otimes \hat{E}^x
\end{equation}
where generically we introduce the light-cone-frame by\footnote{ In order not
  to confuse the reader we define the directional derivatives $\theta_{a}$ to
  be the duals of the frame 1-forms $E^{a}$, {\em i.e.\/} we have
  $E^{a}(\theta_{b})= \delta^{a}{}_{b}$.  We shall reserve the notation
  $\partial_{x}$ for the directional derivative on the base space, namely
  $\partial_{x}\equiv e_{x}{}^{m}\partial_{m}$.  }
\begin{equation}
  \label{eq:KundtCurv1}
  \left\{
  \begin{array}{lclclcl}
    E^{+} & =& du &\hspace{.4cm},\hspace{.4cm}& \theta_{+} & =& \partial_{u} \ -\ H \partial_{v} \; , \\
    E^{-} & =& dv + Hdu + S_{m}dy^{m} & ,& \theta_{-} & =& \partial_{v} \; ,\\
    E^{x} & =& {e_{m}}^{x}\ dy^{m} & ,& \theta_{x} & =& {e_{x}}^{m}\left[ \partial_{m} \ -\ S_{m}\partial_{v}\right] \; ,
  \end{array}\right.
\end{equation}
where the Vielbein on the base space $e_{i}^{x}$ is independent of $v$; the
only $v$-dependence resides in $H$ and $\hat{S}\equiv S_{m}dy^{m}$.
\\
This is the kind of metric that appeared in the characterisations of the null
cases above, eqs.~(\ref{eq:44}) and (\ref{eq:15}), where we defined the
correspondence between the $(n-2)$-bein and the base space metric as
$\mathsf{h}_{mn}\equiv e^i_m e^i_n$.
\\
 
\par
Defining the spin-connection $\omega^{a}{}_{b}\equiv E^{c}\
\omega_{c}{}^{a}{}_{b}$ by means of $dE^{a}=\omega^{a}{}_{b}\wedge E^{b}$ and
imposing it to be metric compatible $\omega_{(ab)}=0$, leads to
\begin{eqnarray}
  \label{eq:KundtCurv2a}
  \omega_{+-} & =& -\theta_{-}H\ E^{+} 
             \, -\ \textstyle{1\over 2}\ \theta_{-}S_{x}\ E^{x} \; ,\\
  \label{eq:KundtCurv2b}
  \omega_{+x} & =& -\left( \theta_{x}H \ -\ e_{x}^{m}\theta_{+}S_{m}\right)\ E^{+}
             \, +\ \textstyle{1\over 2}\theta_{-}S_{x}\ E^{-} 
             \ - \left[\ T_{yx} 
                     + e_{(y}^{m}\theta_{+}e_{x)m}
              \right]\, E^{y}\; ,\\
  \label{eq:KundtCurv2c}
  \omega_{-x} & =& \textstyle{1\over 2}\theta_{-}S_{x}\ E^{+} \; ,\\
  \label{eq:KundtCurv2d}
  \omega_{xy} & =& -\lambda_{zxy}\ E^{z} 
       \ -\  \left[\ T_{xy}
               \ -\ e_{[x}^{m}\theta_{+}e_{y]m}
             \right]\ E^{+} \; , 
\end{eqnarray}
where we defined $\mathsf{d} E^{x} = \lambda^{x}{}_{y}\wedge E^{y}$ and also
defined $\lambda_{zy}=\delta_{zx}\lambda^{x}{}_{y}$, whereas
$\omega_{xy}=\eta_{xz}\omega^{z}{}_{y}$ so that the sign difference is
paramount.\footnote{Observe that a similar condition holds for defining
$e_{mx}=e_{m}{}^{x}$.}
{}Furthermore, we defined
\begin{equation}
  \label{eq:9}
  T_{xy} \; \equiv\; e_{[x}{}^{m}\theta_{y]}S_{m} \hspace{.7cm}\mbox{which for $n=6$ reads:}\hspace{.4cm}
   T_{ij} \ =\ v\ \mathsf{F}_{ij} \; -\; \textstyle{1\over 2}\ \left[ \mathsf{D}\varpi\right]_{ij} \; .
\end{equation}
\par
If we impose that the only $u$-dependency resides in $H$, the non-vanishing
components of the Ricci tensor become
\begin{eqnarray}
  \label{eq:KundtCurv3a}
  R_{++} & =& -\nabla^{(\lambda )}_{x}\partial_{x}H 
           \ +\   \theta_{-}H\ \nabla^{(\lambda )}_{x}S_{x}
           \ -\ H\ \nabla^{(\lambda )}_{x}\theta_{-}S_{x} \nonumber \\
           & & +\ 2S_{x}\ \partial_{x}\theta_{-}H 
            \ -\ \theta_{-}S_{x}\ \partial_{x}H 
            \ -\ S_{x}S_{x}\ \theta_{-}^{2}H \; ,\\
  \label{eq:KundtCurv3b}
  R_{+-} & =& -\theta_{-}^{2}H\ -\ \textstyle{1\over 2}\theta_{-}S_{x}\theta_{-}S_{x} \ +\ \textstyle{1\over 2} \nabla^{(\lambda )}_{x}\ \theta_{-}S_{x} \; ,\\
  \label{eq:KundtCurv3c} 
  R_{+x} & =& -\theta_{x}\theta_{-}H
                     \ -\ \nabla^{(\lambda )}_{y}T_{xy} 
                     \ +\ S_{y}\theta_{-}T_{xy}
                     \ +\ T_{xy}\theta_{-}S_{y} \; , \\
  \label{eq:KundtCurv3d} 
  R_{xy} & =& \mathsf{R}(\lambda )_{xy} \ -\ \nabla^{(\lambda )}_{(x|}\ \theta_{-}S_{|y)} \ +\ \textstyle{1\over 2}\ \theta_{-}S_{x}\theta_{-}S_{y} \; ,
\end{eqnarray}
The Ricci scalar is then given by
\begin{equation}
  \label{eq:KundtCurv4}
  R \; =\; -2\theta_{-}^{2}H
        \ -\ \textstyle{3\over 2}\ \theta_{-}S_{x}\theta_{-}S_{x}
        \ +\ 2\nabla^{(\lambda )}_{x}\theta_{-}S_{x}
        \ -\ \mathsf{R}(\lambda ) \; .
\end{equation}
} 

\end{document}